\documentclass[aps,pra,a4paper,twocolumn,noshowpacs,superscriptaddress,floatfix]{revtex4}
\usepackage{amsfonts}
\usepackage{amsmath}
\usepackage{amssymb}
\usepackage{array}
\usepackage{graphicx}
\usepackage{epstopdf}
\usepackage{multirow}
\usepackage{booktabs}
\usepackage{makecell}
\usepackage{setspace}
\usepackage{booktabs}
\usepackage{threeparttable}
\usepackage[figuresright]{rotating}

\newcommand{\PreserveBackslash}[1]{\let\temp=\\#1\let\\=\temp}
\newcolumntype{C}[1]{>{\PreserveBackslash\centering}p{#1}}
\newcolumntype{R}[1]{>{\PreserveBackslash\raggedleft}p{#1}}
\newcolumntype{L}[1]{>{\PreserveBackslash\raggedright}p{#1}}

\begin{document}

\title{Angular part of trial wavefunction for solving helium Schr\"{o}dinger equation}

\author{{Sanjiang Yang} \footnote{s.yang@whu.edu.cn}}
\affiliation{College of Physics and Electronic Science, Hubei Normal University, Huangshi 435002, China}

\begin{abstract}
In this article, the form of basis set for solving helium Schr\"{o}dinger equation is reinvestigated in perspective of geometry. With the help of theorem proved by Gu $et~al.$, we construct a convenient variational basis set, which emphasizes the geometric characteristics of trial wavefuncions. The main advantage of this basis is that the angular part is complete for natural $L$ states with $L + 1$ terms and for unnatural $L$ states with $L$ terms, where $L$ is the total angular quantum number. Compared with basis sets which contain three Euler angles, this basis is very simple to use. More importantly, this basis is quite easy to be generalized to more particle systems.
\end{abstract}
\maketitle

\section{Introduction}
\label{Introduction}
Helium as one of the simplest three-particle systems always be seen as an ideal model for testing methods to solve Schr\"{o}dinger equations of few-particle systems, e.g. \cite{Modified_Configuration, nakatsuji2000structure, nakatsuji2004scaled}. In variational scheme, since helium consisted of one nucleus and two electrons, after removing the motion of center of mass and separating spin parts, it's quite natural to use the following basis set to search the bound energy levels of system,
\begin{equation}
\begin{aligned}
     \{ |\phi_{ij} \rangle = a_i^{\dagger}a_j^{\dagger}|0 \rangle, \, i,j \le \Omega \}. 
\end{aligned}
\end{equation}
Where $a_i^{\dagger}$ is the creation operator generating a hydrogen-like state $|i \rangle$, $\Omega$ indicates a restriction on space of trial states. For the states of a given total angular quantum number $L$ and total magnetic $M$, the coupled basis $(a_i^{\dagger} \otimes a_j^{\dagger})^L_M$ is more appropriate to use. Noticing that for different configurations, matrix elements of Hamiltonian 
\begin{equation}
\begin{aligned}
     \langle \phi_{ij}^{LM} | H | \phi_{kl}^{LM} \rangle
\end{aligned}
\end{equation}
could be non-zero, ``interaction'' between configurations should be considered carefully in calculations. For more particle systems, the angular coupling scheme is quite complex to build a $|L M \rangle$ trial basis set. Just only for helium, the huge expansion of angular parts makes it difficult to get accurate results for the strong correlated states, e.g. lower lying $S$ states.

For three- or four- particle systems, Hylleraas-type basis sets (HBS) stand out for their high convergence rates \cite{drake2008high, drake2006springer, Yan_Drake, PhysRevLett.113.263007, korobov2002nonrelativistic, PhysRevLett.113.023004, frolov2007field, frolov2012}. HBS are capable of describing the two-electron coalescences, and the contained Hylleraas factors, $r_{ij}^c$, could generate further configurations for basis sets. For helium, Schwartz points out that since angular terms 
\begin{equation}
\begin{aligned}
     \Lambda_{l_1 + 1, l_2 + 1}^{ L M} , \,
     \Lambda_{l_1 + 1, l_2 - 1}^{ L M} , \,
     \Lambda_{l_1 - 1, l_2 + 1}^{ L M} , \,
     \Lambda_{l_1 - 1, l_2 - 1}^{ L M} 
\end{aligned}
\end{equation}
could be generated by 
\begin{equation}
\begin{aligned}
     \vec{r}_1 \cdot \vec{r}_2 \Lambda_{l_1, l_2}^{ L M} ,
\end{aligned}
\end{equation}
where $\Lambda_{l_1, l_2}^{ L M}$ is the vector coupled product of angular momentum $l_1$, $l_2$ for two electrons, the following terms
\begin{equation}
\begin{aligned}
     (l_1,l_2) &= (1,0), (0,1), \\
     (l_1,l_2) &= (1,1)
\end{aligned}
\end{equation}
are enough to calculations on natural or unnatural P states of helium with adequate Hylleraas factors, respectively. Similarly for natural or unnatural D states, the calculations need only three or two terms for angular parts respectively. For general angular momentum case, Schwartz symbolically presented an expression in Cartesian vector form \cite{schwartz1961lamb}. However, this form is inconvenient to use in calculations, especially for unnatural parity states. A standard strategy for choosing angular parts of Hyllerass basis could be found in review \cite{levin2013long}. This strategy advises the terms
\begin{equation}
\begin{aligned}
     (0,L;LM), (1,L-1;LM), \dots ([L/2],L-[L/2];LM)
\end{aligned}
\end{equation}
should be included in calculations on natural parity states. Where $[\quad ]$ denotes ``greatest integer in''. It's quite interesting that this strategy contains no angular parts for S states. Actually this reflects Hylleraas' original insight \cite{hylleraas1929neue, hylleraas1930grundterm}, that is, for S states the wavefuncions of system contain no information of angles, and only depend on distances between particles. This insight is particularly suitable to be expressed in perspective of geometry, i.e., the positions of particles could be determined by a simple rigid body, triangular, with three Euler angles $\alpha$, $\beta$ and $\gamma$. Thus the wavefunctions of helium could be described in the following form
\begin{equation}
\begin{aligned}
     \psi(\alpha,\beta,\gamma,r_1,r_2,r_{12}) .
\end{aligned}
\end{equation}
Here $r_{12}$ could be equivalently replaced with the angle between $\vec{r}_1$ and $\vec{r}_2$, $\theta$. Breit derived radial equations of helium for P states using the latter form \cite{breit1930separation}. According to Wigner \cite{wigner2012group}, the wavefuncions of helium could be labelled by $L$ and $M$, and these states of given $L$ and $M$ satisfied
\begin{equation}
\begin{aligned}
     D(\alpha,\beta,\gamma)\psi_{LM} = \sum_{M^{\prime}} \psi_{LM^{\prime}} D^{L}_{M^{\prime}M}(\alpha,\beta,\gamma)  .
\end{aligned}
\end{equation}
Here we use the convention
\begin{equation}
\begin{aligned}
     D(\alpha,\beta,\gamma) = e^{-\frac{i}{\hbar} \alpha L_z} e^{-\frac{i}{\hbar} \beta L_y} e^{-\frac{i}{\hbar} \gamma L_z} .
\end{aligned}
\end{equation}
  It implies that
\begin{equation}
\begin{aligned}
     \psi _{E,L,M}&(\alpha ,\beta ,\gamma ,R)= \\
     &\sum_{M^{\prime}}{\psi _{E,L,M^{\prime}}(0,0,0,R )D_{M,M^{\prime}}^{L*}}(\alpha ,\beta ,\gamma) .
\end{aligned}
\end{equation}
Where $R$ denotes the internal variables which could describe the rigid body made up with $\vec{r}s$ of particles. For helium, $R$ could be chosen as $r_1$, $r_2$ and $r_{12}$. This formula indicates the calculation needs only $2L + 1$ angular terms for arbitrary $LM$ state. Containing Euler angles makes basis sets a bit complex in calculations \cite{bhatia1964symmetric, pont1995decomposition}. According to the theorem proved by Gu $et~al.$, Wigner D-function $D^{L*}_{LM}$ could be expanded by the so called generalized harmonic polynomials $Q^{L\lambda}_q(\vec{r}_1,\vec{r}_2)$ \cite{gu2001conservation}. For a given $L$, $Q^{L\lambda}_q(\vec{r}_1,\vec{r}_2)$ contains only $2L + 1$ terms, and these terms could be further classified by parity operator. Using $Q^{L\lambda}_q(\vec{r}_1,\vec{r}_2)$, Gu $et~al.$ derived $2L + 1$ coupled generalized radial equations for N-body systems \cite{gu2001independent}. However, the generalized radial equations are not so easy to solve, and the exchange property between identical particles makes generalized radial equations quite complex. In this article, we will use $Q^{L\lambda}_q(\vec{r}_1,\vec{r}_2)$ to construct a variational basis, called geometric basis set (GBS). In calculations on natural parity states of helium, this basis is coincide with Hylleraas basis. For unnatural parity states, angular part of this basis is quite simple to use.

In next section \ref{Geometric basis}, we will briefly construct GBS using $Q^{L\lambda}_q(\vec{r}_1,\vec{r}_2)$. The exchange property between identical particles is easy to deal with since every basis function is definite. A discussion of relations between GBS and HBS is contained in this section as well. Numerical results of energy levels for $n\;^{1,3}P^{e}$ and $n\;^{1,3}D^{o}$ states, with $n$ up to 5, of helium and discussions are presented in Sec.~\ref{Numerical results and Discussion}, together with comparison with other precise results. Finally, a summary is given in Sec.~\ref{Summary}. Atomic units are used throughout.

\section{Geometric basis set}
\label{Geometric basis}

The Schr\"{o}dinger Hamiltonian of helium under assumption of infinitely nucleus mass is 
\begin{equation}
\begin{aligned}
     H=\sum^{2}_{i=1} \left( -\frac{1}{2m} \vec \nabla_i^2 -\frac{Z}{r_i} \right) + \frac{1}{r_{12}}\ .
\end{aligned}
\end{equation}
Where $m$ is the electron mass, $Z$ is the nuclear charge, $r_i$ is the distance between the i-th electron and nucleus, $r_{12}$ is the distance between two electrons. As introduced in Sec.~\ref{Introduction}, we expand trial wavefunctions into GBS,
\begin{equation}
\begin{aligned}
     \phi_{i,j,c,q}^{L \lambda}(\vec{r}_1,\vec{r}_2) = r_1^i r_2^j r_{12}^c &e^{-\alpha_k r_1 -\beta_k r_2 -\gamma_k r_{12}} \times \\
      &Q^{L\lambda}_q(\vec{r}_1,\vec{r}_2) \pm (1 \leftrightarrow 2) \, .
\end{aligned}
\end{equation}
Where 
\begin{equation}
\begin{aligned}
     Q^{L\lambda}_q(\vec{r}_1,\vec{r}_2) = X&^{q-\lambda} Y^{L-q} Z^{\lambda} \, ,	\\
     &\lambda \le q \le L \, , \quad \lambda = 0,1 \, ,
\end{aligned}
\end{equation}
and
\begin{equation}
\begin{aligned}
     X &= x_1 + i y_1 \, , \\
     Y &= x_2 + i y_2 \, , \\
     Z &= z_2X - z_1Y \, .
\end{aligned}
\end{equation}
$i$, $j$ and $c$ are positive integers, restricted by condition
\begin{equation}
\label{restrict}
\begin{aligned}
     i + j + c \le \Omega \, .
\end{aligned}
\end{equation}
$\alpha_k$, $\beta_k$ and $\gamma_k$ are non-linear optimization parameters. $k$ indicates different sectors. $+/-$ is determined by the total spin state of two electrons. The number of angular parts of this basis for a given total angular momentum $L$ and total magnetic $M = L$ is $2L + 1$. The further classification given by parity makes calculations on natural states need only $L + 1$ angular terms and on unnatural states only $L$ tems. The others $M$ states could be obtained by the lower operator $L_-$.

Notice that $X$, $Y$ and $Z$ are identical to $r_1 Y_{11}(\vec{r}_1)$, $r_2 Y_{11}(\vec{r}_2)$ and $r_1 r_2 Y_{11}(\vec{r}_1)Y_{10}(\vec{r}_2) - r_1 r_2 Y_{11}(\vec{r}_2)Y_{10}(\vec{r}_1)$, respectively. And that $M$ equals to its highest value $L$ makes only $r_1^l r_2^{L-l} Y_{l,l}(\vec{r}_1)Y_{L-l,L-l}(\vec{r}_2)$ remains from the angular terms contained by $X^l Y^{L-l}$. This is same to that $\Lambda_{l, L-l}^{ L M}(\vec{r}_1, \vec{r}_2)$ contains only one term $Y_{l,l}(\vec{r}_1)Y_{L-l,L-l}(\vec{r}_2)$ if we set $M=L$. Thus, actually, for $| LL \rangle$ with natural parity states of helium, GBS is coincide with HBS. For unnatural states, GBS provides a clear complete expansion for angular parts, 
\begin{equation}
\begin{aligned}
     Q^{L, 1}_q(\vec{r}_1,\vec{r}_2) = ZQ^{L-1, 0}_{q-1}(\vec{r}_1,\vec{r}_2) \, .
\end{aligned}
\end{equation}

\section{Numerical results and Discussion}
\label{Numerical results and Discussion}

In our calculations two sectors are used. $\alpha_1$ and $\beta_1$ of the first sector are arranged to describe the asymptotic behaviour of wavefunctions. $\alpha_2$ and $\beta_2$ to describe the complex inner correlation effects. Containing of $\gamma$s makes GBS more flexible, and these parameters could be negative in calculations. These six parameters are optimized using Nelder-Mead method \cite{Nelder_Mead}, with initial points around $(1.0,~0.3,~0,~2.5,~2.5,~0)$ for $P$ states and around $(1.0,~0.2,~0,~1.5,~1.5,~0)$ for $D$ states respectively. Under 569 terms parameters of $2\,^3P^e$ state are optimized to around $(1.0650,~1.2681,~0.0546,~3.9895,~2.8728,~0.4738)$. Corresponding result is $E = -0.7105001556783312$, more precise than result of Eiglsperger $et~al.$ $E = -0.7104998$ \cite{eiglsperger2010spectral}, of Kar $et~al.$ $E = -0.7105001556783$ \cite{kar2009one}, and of Hilger $et~al.$ $E = -0.710500155678331$ \cite{hilger1996accurate}. The more detailed parameters values, including of other states, are tabulated in Table~\ref{paras}. These parameters are truncated to 16 digits. As can be seen in this table, the first two parameters $\alpha_1$ and $\beta_1$ for $2\,^3P^e$ state approach to unscreened hydrogenic values $(1.0,~1.0)$. Except for this state, the first two parameters of the rest states approach to screened hydrogenic values $(Z/2,~(Z-1)/n)$, respectively.

Table~\ref{convergence} presents the convergence study of energy level for $2\,^3P^e$ state as scale of basis set $N$ increase. The second column displays energy levels calculated under different $N$. And the third column shows the convergence rates defined as 
\begin{equation}
\begin{aligned}
     R(N_i) = \frac{E(N_{i-1}) - E(N_{i-2})}{E(N_{i}) - E(N_{i-1})} .
\end{aligned}
\end{equation}
It can be seen that the changes of energy level results are quite slow when $N$ greater than 1837. The extrapolated result is obtained at $N_{max} = 4584$, using the following formular
\begin{equation}
\begin{aligned}
     E = E({N_{max}}) + \frac{E({N_{max}}) - E({N_{max}-1})}{R({N_{max}}) - 1} .
\end{aligned}
\end{equation}
The extrapolated value for $2\,^3P^e$ state is $E = -0.71050015567833143120(1)$, which has 5 significant digits more precise than $E = -0.710500155678331$ obtained by Hilger $et~al.$ \cite{hilger1996accurate}. Our calculations provide the most precise nonrelativistic energy level for $2\,^3P^e$ state of helium at present.

The extrapolated values of energy levels for $n\;^{1,3}P^{e}$ and $n\;^{1,3}D^{o}$ states, with $n$ up to 5, of helium are tabulated in Table~\ref{energy_levels}, together with comparison with other available data.
Our results are listed in the second column, and consist with other precise values. Except for $5\;^3P^{e}$ state our results are the most accurate in these calculations. Especially for $2\;^3P^{e}$, $3\;^1P^{e}$ and $3\;^{1,3}D^{o}$ states, our results are accurate to at least 18 significant digits. In our calculations the convergence rates decrease with the principle number increasing. We think it caused by that the relative small scale of basis set limits optimization on non-linear parameters $\alpha_2$, $\beta_2$ and $\gamma$s of basis. The style of restriction on $i$, $j$ and $c$, formular (\ref{restrict}), contains some redundancy when $i$, $j$ or $c$ needs to take a larger number. In calculations on higher exited states, more sophisticated strategy to select $i$, $j$ and $c$ is suggested.

\section{Summary}
\label{Summary}
In this article we reinvestigated the form of basis set for solving helium Schr\"{o}dinger equation and constructed GBS. The angular part of this basis is complete for natural $L$ states with $L + 1$ terms and for unnatural $L$ states with $L$ terms. This basis is quite simple to use. For natural states GBS is coincide with HBS, and for unnatural states this basis provides a clear complete expansion for angular part. Using this basis, we calculated the energy levels for unnatural $n\;^{1,3}P^{e}$ and $n\;^{1,3}D^{o}$ states, with $n$ up to 5, of helium. Our results are accurate to at least 14 significant digits. Except for $5\;^3P^{e}$ state our results provide the most precise values of nonrelativistic energy levels for $3\;^{1}P^{e} - 5\;^{1}P^{e}$, $2\;^{3}P^{e} - 4\;^{3}P^{e}$ and $n\;^{1,3}D^{o}$ states, with $n$ up to 5, of helium up to date. This work shed a light on the reduction of angular parts for more particle systems in variational scheme. The application of GBS to more particle systems is very appealing.

\section{ACKNOWLEDGMENTS}
This work is supported by the National Natural Science Foundation of China (No. 12074295). The numerical calculations in this article have been done on the supercomputing system in the Supercomputing Center of Wuhan University.

\makegapedcells
\setcellgapes{4pt}
\begin{table*}[!htbp]
\caption{Non-linear parameters of $n\;^{1,3}P^{e}$ and $n\;^{1,3}D^{o}$ states, with $n$ up to 5, of helium. The uppers of two $\alpha$s belonging to one state are $\alpha_1$, and the lowers are $\alpha_2$. Values of the third and fourth columns are listed in the same way. Numbers in square brackets indicate power of 10. These parameters are truncated to 16 digits.}
\label{paras}
\begin{tabular}{C{2cm}L{4.5cm}L{4.5cm}L{4.5cm}}
\hline
\hline
States&\qquad $\alpha$s&\qquad $\beta$s&\qquad $\gamma$s \\
\hline
$3\, ^1P^e$&1.020361050922866 &3.232463550080948[--1] &~~3.366471931552636[--1] \\
      &1.560427751553588 &3.672858338184518[--1] &~~2.249793460117225[--1] \\
$4\, ^1P^e$&9.894856701288454[--1] &2.859056951577018[--1] &~~4.038911968792138[--2] \\
      &2.988929095480231[--1] &1.184480106859586 &~~1.519718884659328[--1] \\
$5\, ^1P^e$&1.060255398647911 &2.015091074335774[--1] &~~1.353082000764144[--1] \\
      &3.154503006622944 &9.840808109190531[--2] &~~1.729154147491824[--1] \\
$2\, ^3P^e$&1.065018083101446 &1.268134389751646 &~~5.469290849017121[--2] \\
      &3.989558374309357 &2.872819955241898 &~~4.738929927516252[--1] \\
$3\, ^3P^e$&9.756807688136069[--1] &3.451362765683647[--1] &~~1.046900457212712[--1] \\
      &9.152656935406467[--1] &2.039682069961957 &~~1.381961639113952[--1] \\
$4\, ^3P^e$&9.996417233195907[--1] &3.005244126537764[--1] &~~4.793127476919042[--3] \\
      &3.055517825371942[--1] &1.125278199803706 &~~1.575463294914782[--1] \\
$5\, ^3P^e$&1.395663522693290 &2.478379109498778[--1] &~~8.772884073157730[--2] \\
      &2.934787292316236 &2.598097944716208[--1] &--2.346534637354245[--2] \\
      
$3\, ^1D^o$&1.184022047197198 &3.639115505305711[--1] &~~2.745388851166835[--1] \\
      &5.392132056992792[--1] &1.116396206638743 &~~2.640440963193405[--1] \\
$4\, ^1D^o$&9.780383621596197[--1] &2.638284946020151[--1] &~~2.566333284914318[--1] \\
      &1.157199317324475 &2.780370281045006[--1] &~~8.467593358179396[--2] \\
$5\, ^1D^o$&1.018699271238652 &2.094098615317361[--1] &~~1.455609315523356[--1] \\
      &3.442736030959944 &2.780081768971472[--1] &--7.963177865653946[--2] \\
$3\, ^3D^o$&1.001162390487614 &3.488835451505093[--1] &~~8.489186471007965[--2] \\
      &1.323984041213192 &9.371324188024985[--1] &~~1.466575707090857[--1] \\
$4\, ^3D^o$&1.029172918101772 &2.707198459183539[--1] &~~1.828521293350875[--1] \\
      &3.326973914910362 &2.815195229698139[--1] &--3.387768747560860[--2] \\
$5\, ^3D^o$&1.001931797954439 &1.768993125767199[--1] &~~1.570409312790884[--1] \\
      &2.981338418791743[--1] &9.357083419748404[--1] &~~5.651017549353730[--2] \\
\hline
\hline
\end{tabular}
\end{table*}

\makegapedcells
\setcellgapes{4pt}
\begin{table}[!htbp]
\caption{Convergence study of energy level for $2\,^3P^e$ state of helium as scale of basis set $N$ increase. Number in parentheses is computational uncertainty. Units are a.u.}
\label{convergence}
\begin{tabular}{C{1.5cm}L{4.5cm}L{2.5cm}}
\hline
\hline
$N$  &\qquad E&\quad R \\
\hline
322  & --0.7105001556782 & \\
403  & --0.71050015567832 & \\
497  & --0.710500155678330 & 9.78763 \\
605  & --0.7105001556783313 & 18.1934 \\
728  & --0.71050015567833141 & 5.51508 \\
867  & --0.710500155678331425 & 7.19591 \\
1023 & --0.710500155678331429 & 3.21773 \\
1197 & --0.7105001556783314302 & 3.41477 \\
1390 & --0.7105001556783314307 & 2.28333 \\
1603 & --0.7105001556783314309 & 2.22359 \\
1837 & --0.71050015567833143106 & 1.98654 \\
2093 & --0.71050015567833143112 & 1.88253 \\
2372 & --0.71050015567833143115 & 1.81380 \\
2675 & --0.71050015567833143117 & 1.78975 \\
3003 & --0.71050015567833143118 & 1.73167 \\
3357 & --0.710500155678331431190 & 1.66699 \\
3738 & --0.710500155678331431194 & 1.63022 \\
4147 & --0.710500155678331431197 & 1.62221 \\
4585 & --0.710500155678331431198 & 1.60408 \\
Extrap.& --0.71050015567833143120(1) & \\
\hline
\hline
\end{tabular}
\end{table}

\makegapedcells
\setcellgapes{4pt}
\begin{table*}[!htbp]
\caption{Comparison of nonrelativistic energy levels for $n\;^{1,3}P^{e}$ and $n\;^{1,3}D^{o}$ states, with $n$ up to 5, of helium. Numbers in parentheses are computational uncertainties. Units are a.u.}
\label{energy_levels}
\begin{tabular}{C{1.5cm}L{4.5cm}L{3cm}L{3.5cm}L{4cm}}
\hline
\hline
States&\qquad This work&\qquad Ref.~\cite{eiglsperger2010spectral}&\qquad Ref.~\cite{kar2009doubly}&\qquad References \\
\hline
$3\, ^1P^e$& --0.5802464725943857731(1)& --0.5802464715& --0.580246472594 & --0.580246472594385$^{\rm b}$\\
$4\, ^1P^e$& --0.540041590938513429(1)& --0.5400415905& --0.54004159009  & --0.540041590938513$^{\rm b}$\\
$5\, ^1P^e$& --0.5241789818114119(1)& --0.5241789816& --0.5241790      & --0.524178981811411$^{\rm b}$\\
$2\, ^3P^e$& --0.71050015567833143120(1)& --0.7104998   & --0.7105001556783$^{\rm a}$& --0.710500155678331$^{\rm b}$\\
$3\, ^3P^e$& --0.5678128987251561(1)& --0.56781281  & --0.567812898724 & --0.567812898725152$^{\rm b}$\\
$4\, ^3P^e$& --0.535867188768223(2)& --0.53586715  & --0.5358671887   & --0.535867188768211$^{\rm b}$\\
$5\, ^3P^e$& --0.52225457570724(2)& --0.52225456  & --0.52225457     & --0.522254575707233$^{\rm b}$\\
      
$3\, ^1D^o$& --0.563800420462367542(2)& --0.563800418 & --0.563800420462 & --0.56380042$^{\rm c}$\\
$4\, ^1D^o$& --0.5345763855561833(1)& --0.5345763848& --0.534576385556 & --0.53457638$^{\rm c}$\\
$5\, ^1D^o$& --0.521659015466304(2)& --0.5216590151& --0.521659015466 & --0.52165901$^{\rm c}$\\
$3\, ^3D^o$& --0.559328263097247843(1)& --0.55932824  & --0.559328263096 & --0.55932826$^{\rm c}$\\
$4\, ^3D^o$& --0.532678601895944(1)& --0.55932824  & --0.532678601895 & --0.53267860$^{\rm c}$\\
$5\, ^3D^o$& --0.52070346202795(1)& --0.520703455 & --0.520703462028 & --0.52070345$^{\rm c}$\\
\hline
\hline
\end{tabular}
\begin{tablenotes}
	\footnotesize
	\item[1] $^{\rm a}$ Referece~\cite{kar2009one}
	\item[2] $^{\rm b}$ Referece~\cite{hilger1996accurate}
	\item[3] $^{\rm c}$ Referece~\cite{PhysRevA.78.032505}
\end{tablenotes}
\end{table*}

\bibliography{Geometric_basis}

\end{document}